\documentclass[11pt,a4paper]{article}
\usepackage[utf8]{inputenc}
\usepackage[T1]{fontenc}
\usepackage{lmodern}
\usepackage{microtype}
\usepackage[margin=2.cm]{geometry}
\usepackage{setspace}
\usepackage{parskip}
\usepackage{graphicx}
\usepackage{booktabs}
\usepackage{enumitem}
\usepackage{hyperref}
\hypersetup{colorlinks=true,linkcolor=blue,citecolor=blue,urlcolor=blue}
\usepackage{titling}
\usepackage{fancyhdr}
\usepackage{caption}
\usepackage{amsmath,amssymb}

\usepackage[medium]{titlesec}

\usepackage[symbol]{footmisc}
\usepackage[round, comma, authoryear]{natbib}
\pretitle{\vspace*{-1cm}\begin{center}\LARGE\bfseries}
\posttitle{\par\end{center}\vskip 0.5em}
\preauthor{\begin{center}\large}
\postauthor{\end{center}}
\predate{\begin{center}\large}
\postdate{\end{center}}

\let\OLDthebibliography\thebibliography
\renewcommand\thebibliography[1]{
  \OLDthebibliography{#1}
  \setlength{\parskip}{0pt}
  \setlength{\itemsep}{0pt plus 0.3ex}
}

\usepackage{multicol}
\fancypagestyle{titlepageheader}{
  \fancyhf{} 
  \fancyhead[L]{ESO White Paper Call: Expanding horizons}
  
}

\renewcommand{\footnoterule}{
  \kern -3pt
  \hrule width \textwidth
  \kern 2.6pt
}

\begin{document}


\begin{titlepage}
  \thispagestyle{titlepageheader}
  {\LARGE \textbf{Expanding stellar horizons with polarized light}\par}
  
  \vspace{0.2cm}

  \textbf{Authors:} J.\,Vandersnickt$^1$\footnote[1]{Corresponding author: jelle.vandersnickt@kuleuven.be}, R.\,Ochoa\,Armenta$^1$, V.\,Vanlaer$^1$, A.\,David-Uraz$^2$, 
  C.\,Aerts$^1$,
  J.-C. Bouret$^3$,
  D.\,M.\,Bowman$^4$,
  L.\,Bugnet$^5$,
  S.\,B. Das$^6$,
  V.\,Khalack$^7$,
  J.\,Labadie-Bartz$^8$,
  S.\,Mathis$^9$, 
  Y.\,Naz\'e$^{10}$,
  C.\,Neiner$^{11}$,
  P.\,Petit$^{12}$,
  V.\,Petit$^{13}$, 
  K.\,Thomson-Paressant$^4$,
  T.\,Van Doorsselaere$^{14}$,
  M.\,Vanrespaille$^1$

{\footnotesize  \setlength{\parskip}{0pt} \setlength{\itemsep}{-3pt}
$^1$ Institute of Astronomy, KU Leuven, Celestijnenlaan 200D, 3001, Leuven, Belgium 

$^2$ Department of Physics, Central Michigan University, Mt. Pleasant, MI 48859, USA

$^3$ Aix-Marseille Univ., CNRS, CNES, LAM, 13388 Marseille, France

$^4$ School of Mathematics, Statistics and Physics, Newcastle University, Newcastle upon Tyne, NE1 7RU, UK

$^5$ Institute of Science and Technology Austria (ISTA), Am Campus 1, 3400 Klosterneuburg, Austria 

$^6$ Center for Astrophysics | Harvard \& Smithsonian, 60 Garden Street, Cambridge, MA 02138, USA

$^7$ Universit\'e de Moncton, Moncton, N.-B. E1A 3E9, Canada

$^8$ DTU Space, Technical University of Denmark, Elektrovej 327, 2800 Kgs., Lyngby, Denmark

$^9$ Universit\'e Paris-Saclay, Universit\'e Paris Cité, CEA, CNRS, AIM, F-91191 Gif-sur-Yvette, France

$^{10}$ Groupe d’Astrophysique des Hautes Energies, STAR, Université de Liège, B5c, Allée du 6 Août 19c, 4000 Sart Tilman, Liège,
Belgium

$^{11}$ LIRA, Observatoire de Paris, Universit\'e PSL, CNRS, Sorbonne Universit\'e, Universit\'e Paris Cit\'e, CY Cergy Paris Universit\'e,
92190 Meudon, France

$^{12}$ Universit\'e de Toulouse; UPS-OMP; IRAP; Toulouse, France

$^{13}$ Department of Physics and Astronomy, Bartol Research Institute, University of Delaware, Newark, DE, 19713, USA

$^{14}$ Centre for mathematical Plasma Astrophysics, Mathematics Department, KU Leuven, Belgium
}

  
  \vspace{0cm}
\centering
\begin{minipage}{0.9\textwidth}
    \centering
     \includegraphics[height=5.3cm]{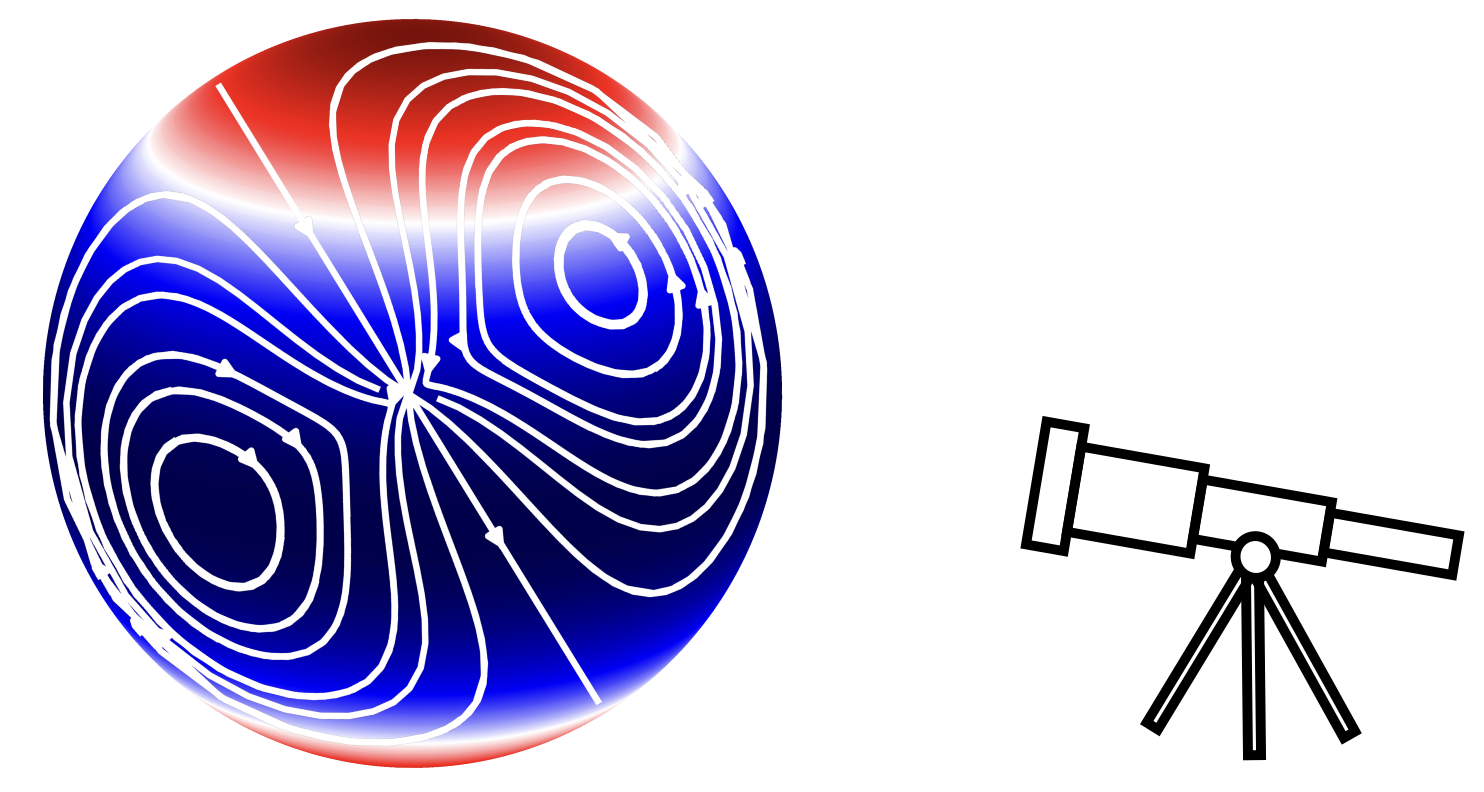}\\
     \small
     Illustration showing an $l=2, m=0$ pulsation mode overlaid with an inclined dipole magnetic field. \\
     Figure inspired by \href{https://4d-star.org/?p=790}{the press release} of \citet{Vandersnickt2025}
\end{minipage}

\vfill

  \centering
  \textbf{Abstract}
  
  \begin{minipage}{0.9\textwidth}
    \small
    The polarization of light is a critically under-utilized, rich source of information 
    in 
    astronomy. 
    For stars in particular, surface magnetism 
    can be detected and measured 
    with spectro-polarimetry
    .
    Many questions about these surface fields remain unanswered due to a lack of dedicated instruments capable of probing weak and strong surface magnetic fields 
    for the entire mass range of stars, 
    from M-dwarfs (and even substellar objects) to massive O-type stars at different evolutionary stages and metallicities.
    These questions range from the origin of these fields to their true incidence rate throughout the stellar population and the dependence on metallicity.
    Magnetic fields, although currently often excluded 
    from stellar evolution models, play an important role in stellar evolution.
    Connecting the surface fields to internal fields through asteroseismology will instigate a new era of understanding stellar evolution 
    and the transport of angular momentum and chemical elements throughout stellar interiors, also impacting our understanding of star--planet interactions and stellar remnants. 
    Polarimetry is also an under-utilized tool to observationally constrain the mode identification of nonradial oscillations, which lies at the basis of accurate asteroseismic parameter estimation at percentage-level for stellar radii, masses, ages, internal rotation, and magnetic field strengths.
    Combining strong constraints on mode identification and surface magnetic properties through the acquisition of time-resolved, high-resolution  and high-signal-to-noise (S/N) spectropolarimetry and spectroscopy promises to bring leaps forward in our understanding of stellar structure, particularly when combined with long-term space photometric data from past, current, and future missions. 
  \end{minipage}

\end{titlepage}

\section{Surface magnetic fields across the HR diagram}\vspace{-11pt}
Three important parameters that determine the evolution of stars across the Hertzsprung-Russell diagram over the course of their lifetimes are mass, rotation, and magnetism, 
with the effects of the latter 
perhaps being the least 
understood
. In the case of high-mass stars, the strong mass loss due to radiation-driven winds is also a crucial phenomenon. For the subset of massive stars ($\lesssim 10 \%$) that exhibit strong large-scale surface magnetism, the fields interact with both mass loss (e.g., \citealt{ud-Doula2002}) and rotation (e.g., \citealt{ud-Doula2009}), with these interactions strongly impacting stellar evolution \citep{Keszthelyi2022}. Any accurate numerical implementation of these effects in evolutionary codes
demands a realistic 
angular momentum transport scheme, which can only be 
properly calibrated using 
the combination of spectro-polarimetry and asteroseismology (see next section). More generally however, the origin of surface magnetic fields in \textit{stars with radiative envelopes} and their relatively flat incidence rate across two orders of magnitude of stellar masses remain poorly understood. \textbf{This all points to large gaps in our present understanding of 
the formation, evolution, and lives of stars, 
particularly concerning the role of magnetic fields.}

While stellar mergers have been proposed as a viable channel to form some magnetic massive stars (e.g., \citealt{Schneider2019}), it has also been suggested that a bistability mechanism \citep{Jermyn2020} is involved in setting the fraction of massive stars with strong magnetic fields. This proposed mechanism states that either a field is strong enough to stabilize the sub-surface convection zone 
\citep{Cantiello2009} and/or inhibit differential rotation (e.g., \citealt{Auriere2007}) and survives, or it 
is ultimately 
destroyed, giving rise instead to 
weak, small-scale 
fields \citep{Bekki2025}. 
Stable weak fields have been observed in a handful of A/Am stars (e.g., \citealt{Blazere2016}), although they are very difficult to detect and characterize with current instrumentation. It is currently unknown whether metallicity might play a role in this proposed bistability mechanism.
\textbf{
Contraining this phenomenon observationally 
by probing whether the incidence of strong surface magnetic fields on high- and intermediate-mass stars varies between environments with different metallicities would significantly contribute to our understanding of various populations of stars throughout the history of our Galaxy and the Universe.
} 

In lower-mass stars, the origin of surface magnetism is somewhat better understood, though several challenges remain. The dominant aspect is the generation of a dynamo field from the turbulent, differentially-rotating convective stellar envelope.  As a result of the strong interplay between rotation and magnetism, models of magnetic rotational spin-down allow age determinations in a process called \textit{gyrochronology} (e.g., \citealt{Barnes2007}). This in turn proves useful in the study of Galactic archaeology by tracing populations with different ages, corresponding to various generations of stellar formation. Nevertheless, important questions remain as the dynamo of Sun-like stars appears to decrease in efficiency for stars more evolved than our Sun, leading to a breakdown in previously determined age-period relationships, and the observation of unexpected, rapidly-rotating old stars \citep{Metcalfe2025}. Moreover, the discovery of slowly-rotating M-dwarfs with strong magnetic fields challenges the weakening of the magnetic field expected from spin-down \citep{Lehmann2024}. Characterizing the magnetic fields of brown dwarfs \citep{Berdyugina2017}, although challenging due to their faintness, constitutes an important step towards understanding the magnetic properties of extrasolar planets, together with the study of magnetic activity in exoplanet hosts, which guide star-planet interactions \citep{strugarek2021, Vidotto2025}. Thus, \textbf{sensitive magnetic measurements of low-mass stars are crucial to characterize the magnetic environments, and even the habitability (with respect to stellar activity), of 
worlds beyond our own}. 

We conclude that the study of magnetism across the entire Hertzsprung-Russell diagram plays a dominant role in fulfilling three of the five ``areas of future scientific discovery'' beyond 2035 that were identified in the \textit{Astronet Roadmap 2022-2035}: (1) the formation of planets, stars, and galaxies; (2) the first stars, galaxies, and the epoch of reionization; and (3) the origins of our Solar system and the (atmospheric) characterization of other worlds.

These proposed science cases presently 
lie beyond the realm of what is currently feasible
, as existing efforts are largely \textit{magnitude-limited}, with current high-resolution spectropolarimeters -- operating either in the visible or near-infrared -- being mounted on telescopes with apertures smaller than $\sim$8\,m.
{\bf Having a high-resolution ($R \geq 100,000$), high-efficiency near-ultraviolet-to-near-infrared spectropolarimeter 
mounted on a telescope with a large ($\geq 10$\,m) aperture will be fundamental to answer the many open questions in the field of stellar magnetism}, which are very likely to remain well into the 2040s, 
as this will allow for the detection of both weak, localized fields on relatively bright stars and strong, large-scale fields on distant (e.g., beyond the Milky Way, something that is currently inaccessible; \citealt{Bagnulo2020}) or intrinsically dim stars.
The wide wavelength range ensures access to the magnetic surfaces at all masses and evolutionary stages. 
\vspace{-15pt}
\section{Probing magnetic fields below the surface of stars}
\vspace{-11pt}
Asteroseismology interprets the oscillations of stars to study their internal structure throughout the Hertzsprung-Russell diagram \citep{Aerts2021}.
Its unique approach probes the internal structure rather than solely the surface properties. 
State-of-the-art asteroseismic techniques compare oscillations predicted by stellar models with observed frequencies to determine both internal and global properties of 
a star with unprecedented precision. 
Since about a decade, space asteroseismology delivers measurements of the internal rotation profiles of stars. This resulted in corrections of theoretical differential rotation profiles with a factor up to 100 across a wide mass range
\citep{Aerts2019}.
Aside from differential rotation,  efficient angular momentum transport through an internal magnetic field has emerged as a leading candidate to explain this discrepancy \citep[e.g.][]{Takahashi2021, Moyano2023, Moyano2024}. 
Over the past few years, such internal fields were detected  in red giants \citep{ Li2022, Li2023, Deheuvels2023, Hatt2024} and in intermediate and massive stars \citep{Vandersnickt2025, Takata2025}.
With additional evidence for internal magnetic fields in even more stars \citep[e.g.][]{Stello2016, Rui2024}, 
{\color{blue}such} fields are expected to be prevalent from low-mass M-dwarfs to massive O-stars. 
With strong magnetic fields detected on the surface of white dwarfs and other stellar remnants, investigating stellar magnetism at every evolutionary stage has become increasingly relevant.

\textbf{A fundamental aspect in obtaining accurate asteroseismic results is identifying the nonradial oscillation modes} by their spherical harmonic indices: the angular degree $l$ and azimuthal order $m$, which characterize the mode's  horizontal structure. 
This is currently obtained through, for instance, pattern recognition from \'echelle diagrams for pressure modes or period spacing patterns for gravity modes \citep{Aerts2021}.
These methods are not available for every type of pulsator 
and break down for stars with significant rotation \citep{Mombarg2025}.
Time-resolved, linear polarimetry in different pass-bands is capable of \textbf{observationally deducing the angular degree $l$ and azimuthal order $m$ of nonradial modes} by comparing the phase-resolved ratio of polarimetric amplitude to photometric amplitude
induced by the geometric distortions and local temperature variations.
In their proof-of-concept study, \citet{Cotton2022} showed that time-resolved polarimetry coupled to space photometry can constrain both the angular degree and the inclination angle between the rotation axis and the line of sight.
The currently available polarimeters are limited to the brightest stars in the sky. 
{\bf A next-generation instrument offering dedicated time-resolved polarimetry for large samples of bright and faint seismic targets would fill a critical observational gap, enabling secure mode identification, which is the pillar of successful asteroseismology.}
 The instrument would have to be capable of resolving a wide range of frequencies for both bright and faint targets with a precision between 1 and 10 part-per-million for an integration time on the order of minutes for targets up to magnitude 12 in the visible band.
 This will then be combined with the space-based photometric data of for instance CoRoT, Kepler, TESS, and  -- by that time -- PLATO and future photometric space missions.  

Detection of pulsation modes through photometric time series of stars is mostly limited to angular degrees up to three, as modes with a higher degree suffer from strong cancellation when observed through photometry. 
This is where time-resolved, high-resolution, and high S/N spectroscopy plays a major role as nonradial oscillations vary the measured line profiles for each angular degree. 
Multi-object, high-resolution ($R \geq 100,000$), large-aperture spectroscopy
with a cadence of minutes will give access to line-profile variations of modes with higher angular degrees ($l \geq 3$) \citep[][Chapter\,6]{Aerts2010}.
This allows to access more probing regions for different pulsation modes than those from space photometry, putting stronger constraints on the internal rotation and magnetic profiles. 
Combining time-resolved spectroscopy with the mode identification and inclination angle constraints of polarimetry advances the determination of both surface and internal rotational and magnetic properties through high-degree modes. 
\textbf{A dedicated instrument based on the combined capabilities of (spectro-)polarimetry and broadband spectroscopy additionally capable of tracking single and multiple stars over long time bases of a few years to investigate the modes on these time scales will open unprecedented avenues for breakthroughs in stellar modelling.}
\vspace{-15pt}
\section{Expanding polarized horizons}
\vspace{-11pt}
The measurements of internal rotation and magnetic fields are limited to layers that identified oscillation modes are sensitive to.
\textbf{Fitting these internal measurements with observationally obtained values on the stellar surface would allow for self-consistent, internal modelling of both rotation and internal magnetic fields from deep in the interior to the surface}, as shown possible by \citet{Lecaonet2022} for the magnetic field. 
This would be fundamental for understanding how core fields influence nonradial oscillations and how these fields couple to small- and large-scale surface fields that drive angular momentum transport and loss.
These highly relevant results are only obtainable through a synergy of theory, modelling, and carefully interpreted observations. 
The coupling of internal and surface magnetic fields will be critical to calibrate 3D stellar structure and evolution models of rotating magnetic stars across all mass regimes.

A joint observational effort combining time-resolved polarimetric measurements with high-resolution spectro-polarimetry and spectroscopy will set forth the future of stellar astrophysics with unprecedented advances in the theory of stellar structure and evolution for single and multiple stars across the Hertzsprung-Russell diagram, allowing us to \textbf{describe magnetic fields, rotation, chemical mixing, and other transport processes from the core to the surface, giving us critical insights in stellar processes we will not be able to explore with other space- or ground-based observatories}. 
We argue for a dedicated instrument that is able to perform simultaneous high-cadence, high-resolution spectroscopy, and  (spectro-)polarimetry, while fully exploiting high-precision space photometry for magnetic and asteroseismic analysis.
Ideally, this instrument would cover parts of the sky where current and future missions such as TESS and PLATO were able to obtain long light curves of seismically active stars, leading the Southern Hemisphere to be the most promising candidate. 

\vfill
\vspace{-10pt}
\begin{multicols}{1}
\scriptsize
\singlespacing 
\itemsep-2em 
\bibliography{refs}  
\end{multicols}

\end{document}